\documentclass[a4paper,11pt]{article}
\usepackage{latexsym}
\usepackage{epsfig, subfigure, subfloat, graphicx, float }
\usepackage{amssymb}
\usepackage{amsmath}
\usepackage{epstopdf}
\usepackage{authblk}
\title{\bf Symmetry breaking and the onset of cosmic acceleration in scalar field models}

\author[1]{H. Mohseni Sadjadi\thanks{mohsenisad@ut.ac.ir}}
\author[2]{M. Honardoost \thanks{m\_honardoost@sbu.ac.ir}}
\author[2]{H. R. Sepangi \thanks{hr-sepangi@sbu.ac.ir}}
\affil[1]{Department of Physics, University of Tehran}
\affil[2]{Department of Physics, Shahid Beheshti University}

\begin{document}

\maketitle

\begin{abstract}
We propose a new scenario for the onset of positive acceleration of our Universe based on symmetry breaking in coupled dark energy scalar field model.
In a symmetry breaking process where the scalar field rolls down its own potential, the potential reduction is not in favor of acceleration.
In our model, when dark matter density becomes less than a critical value, the  shape of the effective potential is changed and, the quintessence field climbs up along {\it{its own potential}} while rolls down the effective potential.  We show that this procedure may establish the positivity of the potential required for the Universe to accelerate. In addition, we show that by choosing an appropriate interaction between dark sectors there is the possibility that the scalar field resides in a new vacuum giving rise to a positive cosmological constant which is responsible for a permanent late time acceleration.

\end{abstract}

\section{Introduction}
To explain the positive acceleration of the Universe in the present era many models have been introduced. The most simple and natural candidate for dark energy is a cosmological constant which was first introduced by Einstein to oppose gravitational attraction resulting in a static universe. This idea was abandoned after the Hubble discovery but resurfaced again later to describe the present acceleration of the Universe \cite{cc}.  The cosmological constant compatible with observation is nearly 120 orders of magnitude less than what is obtained by considering  vacuum energy in the standard particle physics framework with a cutoff at the planck scale. Hence it is reasonable to consider it as a constant of nature \cite{sah}.
In the cosmological constant model dark energy density is constant and the equation of state parameter of dark energy is always -1.  This situation changes when dynamical dark energies are employed.

A natural and simple model for dynamical dark energy is the scalar field model \cite{quin}. In this type of models, the acceleration of the Universe  depends on the potential considered for the scalar field. This is similar to scalar field models of the early universe when inflation occurs during slow roll for nearly flat potentials or  during rapid oscillations for specific power law potentials \cite{rapid}. Note that, due to the presence of dark and ordinary matter at late times the problem is not as straightforward as in the inflationary epoch \cite{shib}. An important specification of scalar field models is their usability in the phenomenon of symmetry breaking in particle physics, where the initial symmetric state becomes unstable and the system rolls down to a new vacuum, breaking the initial symmetry \cite{dj}. Through this formalism and in the context of dark energy models, when the system leaves its false vacuum and settles down to a true vacuum, the cosmological constant is reduced. This formalism with fine tuning of the initial conditions can be considered as a bridge connecting the large value of the cosmological constant in the early universe to its small value at late times \cite{sah}.

In recent years some attempts have been made to relate the onset of the acceleration of the Universe to symmetry breaking in scalar field models.
Inspired by \cite{lind1} where the symmetry breaking were used to end the slow roll inflation, a hybrid quintessence model was introduced in \cite{hibq}. In this model beside the dark energy scalar field, another scalar field is present whose evolution causes the symmetry to brake. In this framework, the Universe may experience an acceleration while the scalar field is rolling down the total potential and also when it settles to the new minimum, provided that an appropriate potential is chosen. The same model was used in \cite{hibdar} to describe the phantom divide line crossing. In \cite{symmetron}, due to the special form of the coupling of matter and scalar field dark energy, an effective potential has been obtained such that when the matter energy density becomes less than a critical value (determined by the parameters of the model) the symmetry is broken; the scalar fields rolls down towards the minimum of the effective potential and acceleration begins. In this model, dubbed symmetron, the matter density plays the role of the trigger field in the hybrid model. The same model has also been used to study the onset of inflation \cite{syminf}. In these models, the fields rolls down along its own potential as well as the effective potential at the same time. So if initially (before the symmetry breaking) the field resides at the extremum of the unbroken effective potential, the symmetry breaking which decreases the value of the potential and increases the kinetic energy, is not in favor of the positive acceleration. This issue will be  explained more in the third section. The acceleration obtained via numerical methods in these models arises from a positive constant term which was initially included (implicitly or explicitly) in the potential and has nothing to do with the symmetry breaking.  Indeed in these types of models the symmetry breaking may be rather used to end the acceleration or the inflation \cite{lind1}.

 In this paper we introduce a new proposal for the onset of  acceleration based on symmetry breaking in a scalar field dark energy model in a spatially flat Friedmann Robertson Walker space time. In our model, after the effective potential shape is changed, contrary  to symmetron and  hybrid models, the quintessence  climbs over its own potential. This procedure may establish the positivity of the potential which is necessary for the acceleration of the Universe.

The organization of the paper is as follows:
 In the second section we derives the role of the scalar field potential in driving the acceleration of the Universe. Our arguments upon which we construct our model is based on this section. In the third section we briefly review two quintessence models in which the symmetry breaking forces the scalar field to descend its own potential: the hybrid and  symmetron models.

 In  section four we introduce a new proposal for late time acceleration based on  symmetry breaking where the scalar field climbs over its own potential while descending down along the effective potential after the effective potential shape is changed into the form dictated by symmetry breaking. To do so, we need to consider an appropriate interaction between the dark sectors which is not linear in terms of dark matter energy density.  In this context, two frameworks are presented, one in which the form of the interaction between dark sectors is borrowed from the scalar-tensor theories  but employs two dark energy scalar fields giving rise to a transient acceleration
 and the other where we  try to obtain a permanent acceleration arisen from symmetry breaking by employing appropriate interactions (not derived from an action)  between dark sectors of the Universe. In this case the Universe tends to a de Sitter space time with positive acceleration. We also
 examine the stability of the model and illustrate our results via numerical depictions.  Our  motivation to propose this model, as stated before, is that in a symmetry breaking procedure where the scalar field rolls down its  potential, as we will show, the potential reduction not only is not in favor of acceleration but also is against it.

{\it{In our study,  by  ``symmetry breaking,''  we generally mean the procedure where the potential gets the shape of a ``symmetry breaking potential'' even though the field has not yet been settled at the new minimum }}.

We use the units $\hbar=c=1$.

\section{Role of positive potential in driving the acceleration }

In a spatially flat Friedmann Robertson Walker space-time, filled (nearly) with dark matter $\rho$ and  dark energy scalar fields $\phi_i$ with potentials $V(\phi_i)$, the Friedmann equations are
\begin{equation}\label{1}
H^2={1\over 3M_P^2}\left({1\over 2}\sum_i\dot{\phi_i}^2+V(\phi_i)+\rho\right),
\end{equation}
\begin{equation}\label{2}
\dot{H}=-{1\over 2M_P^2}\left(\sum_i\dot{\phi}^2+\rho\right),
\end{equation}
where in terms of the scale factor $a(t)$, the Hubble parameter is given by $H={\dot{a}(t)\over a(t)}$. A dot indicates time derivative and   $M_P=2.4\times 10^{18}GeV$ is the reduced Planck mass.
The positive acceleration of the Universe is specified by  $\ddot{a}>0$ which, by using
$\dot{H}+H^2={\ddot{a}\over a^2}$, leads to
\begin{equation}\label{3}
\dot{H}+H^2={1\over 6M_P^2}\left(-2\sum_i\dot{\phi_i}^2+2V(\phi_i)-\rho\right)>0.
\end{equation}
This can be rewritten in terms of the deceleration parameter $q$ as
\begin{equation}\label{4}
q=-{\dot{H}+H^2\over H^2}={1\over 2}(1+3\omega)<0,
\end{equation}
where $\omega$ is the equation of state parameter of the Universe. From (\ref{3}), it is clear that to have a positive acceleration we need a positive potential which drives  acceleration
\begin{equation}\label{5}
V(\phi_i)>\sum_i\dot{\phi_i}^2+{\rho\over 2}.
\end{equation}

In models trying to describe onset of acceleration via symmetry breaking, the scalar field is initially settled down at a fixed point, $\dot{\phi}=0$, then after the symmetry breaking it rolls down its potential. Therefore in theses models, after the symmetry breaking, the potential decreases while the kinetic energy increases and these make $\dot{H}+H^2$ more negative and so are not in favor of acceleration as can be seen from (\ref{3}). In the following section we review two of these models and afterwards we will propose a model which can associate the acceleration to symmetry breaking in a concrete way.

\section {A brief review and critical analysis of models ascribing acceleration to symmetry breaking}

In this part we study the possible relation between the symmetry breaking and the positive acceleration of the Universe in two well known scalar field models.
\subsection{Hybrid quintessence}
Based on the ``hybrid inflation'' model introduced in \cite{lind1}, the Authors of \cite{hibq}  suggested a hybrid quintessence model to study the present acceleration of the Universe.
In this model an additional scalar field $\psi$ is employed to trigger the slow roll of the quintessence field $\phi$ via symmetry breaking at late times.
The potential is taken as
\begin{equation}\label{6}
V(\phi,\psi)=\beta\phi^4+\alpha\phi^2+h\psi^2\phi^2+\lambda\psi^4+\mu\psi^2,
\end{equation}
where $\alpha$, $\beta$, $h$, $\lambda$ and $\mu$ are real constants satisfying
$\{\beta>0,\,\,\, \lambda>0,\,\,\, \mu<0,\,\,\, \alpha>0,\,\,\, h<0\}$ and  $\{h^2-4\lambda\beta<0,\,\,\,  h\mu-2\alpha\lambda>0,\,\,\,  h\alpha-2\beta\mu>0\}$ \cite{hibq}.
When $\psi^2<\psi_c^2$, where $\psi_c^2={-{\alpha\over h}} $, the effective mass squared of $\phi$ is positive and this field is settled down to  $\phi=0$. But when $\psi$ becomes greater than the critical value, $\psi^2>\psi_c^2$, the squared effective mass of $\phi$ becomes negative.  $\phi=0$ as the local maximum of the potential becomes an unstable state, hence $\phi$ rolls down the potential,  leading to the present acceleration of the Universe as claimed in the scenario described in \cite{hibq}. Finally these fields settle down to the minimum of the potential and acquire their final values $\psi_f$, $\phi_f$. Using  $\psi_f^2=-{\mu+h\phi_f^2\over 2\lambda}$, and $\phi_f^2=-{\alpha+h\psi_f^2\over 2\beta}$, we find
\begin{eqnarray}\label{7}
\psi_f^2&=&-{h\alpha-2\beta\mu\over h^2-4\lambda\beta},\nonumber \\
\phi_f^2&=&-{-2\alpha\lambda+\mu h\over h^2-4\lambda \beta}.
\end{eqnarray}
Note that
\begin{equation}\label{8}
V_f(\phi,\psi)={\beta\mu^2+\alpha^2\lambda-\alpha\mu h\over h^2-4\beta \lambda},
\end{equation}
is negative. At the critical value $\psi_c$, the potential is  $V_c=-{\mu \alpha\over h}+{\lambda\alpha^2\over h^2}$, which can be shown to be greater than $V_f$.
So the potential decreases after symmetry breaking as is expected because the fields roll down to the  minimum of the potential.  The symmetry breaking thus reduces the potential and, if the fields were initially at rest,  it is seen that the symmetry breaking is not in favor of acceleration, see (\ref{3}). As the final potential is also negative, to obtain a positive potential required for late time acceleration we must add a  positive cosmological constant $V_0>0$ to the potential;  $V\to V+V_0$. However the acceleration in this model, if happens, as was shown numerically in \cite{hibq}, is driven by the cosmological constant $V_0$ which is not arisen from the symmetry breaking. This is similar to the hybrid inflation model where the acceleration is driven by the vacuum energy density \cite{lind1}.
\subsection{The symmetron model}

Another model used to relate symmetry breaking to the present acceleration of the Universe is the symmetron model presented in \cite{symmetron}. This model uses the scalar tensor
action \cite{khoury}
\begin{equation}\label{9}
S=\int d^4x\sqrt{-g}\left(\frac{M_P^2}{2}R-{1\over 2}g^{\mu \nu}\partial_\mu \phi \partial_\nu \phi-V(\phi)\right)+S_m[\tilde{g^i}_{\mu \nu},\psi^i],
\end{equation}
where $\phi$ is a scalar field and $S_m$ is the action for the $\psi^i$ which can represent dark matter and other species \cite{khoury}. The coupling between the scalar field and each ingredient, $\psi^i$, is realized via $\tilde{g^i}_{\mu \nu}=A_i^2(\phi)g_{\mu \nu}$, where $A_i(\phi)$ is a positive function. As we are interested only in cosmic acceleration resulting from interaction between dark sectors, we assume that $A_i$ is nontrivial $A_i\neq 1$ only for dark species and do not consider ordinary matter and its possible coupling to dark energy. The scalar field equation of motion is derived from ${\delta S\over \delta \phi}=0$
\begin{equation}\label{10}
\ddot{\phi}+3H\dot{\phi}+V_{,\phi}+A^{-1}A_{,\phi}\rho=0.
\end{equation}
The continuity equation for cold dark matter which we consider as a perfect fluid is
\begin{equation}\label{11}
\dot{\rho}+3H\rho=A^{-1}A_{,\phi}\rho\dot{\phi},
\end{equation}
where $\rho$ is dark matter energy density. In terms of re-scaled energy density $\hat{\rho}$ defined by $\rho=A\hat{\rho}$, the above equations may be expressed in a more simple form
\begin{eqnarray}\label{12}
&&\ddot{\phi}+3H\dot{\phi}+V_{,\phi}+A_{,\phi}\hat{\rho}=0,\nonumber \\
&&\dot{\hat{\rho}}+3H\hat{\rho}=0.
\end{eqnarray}
Note that $\hat{\rho}$ has a simple form $\hat{\rho}(a)=\hat{\rho}(a=1)a^{-3}$. The physical density appearing in the Friedmann equations (\ref{1}), (\ref{2}) and (\ref{3}) in the Einstein frame is $\rho=A(\phi)\hat{\rho}$ \cite{symmetron}. The first equation in (\ref{12}) implies that one may consider a ``$\hat{\rho}$ dependent'' effective potential, $V^{eff}_{,\phi}=V_{,\phi}+A_{,\phi}\hat{\rho}$, for $\phi$.  The Friedmann equation obtained from  Einstein equations are
\begin{eqnarray}\label{13}
&&H^2={1\over 3M_P^2}\left({1\over 2}\dot{\phi}^2+V+A\hat{\rho}\right),\nonumber \\
&&\dot{H}=-{1\over 2M_P^2}\left(\dot{\phi}^2+A\hat{\rho}\right).
\end{eqnarray}
Now, with these preliminaries let us study how the model works. It is clear from (\ref{12}) that $\hat{\rho}$ is a decreasing function of time. One can arrange the model such that when $\hat{\rho}<\rho_c$, the symmetry is broken and the scalar field gains a negative mass squared so that the system becomes unstable and the field slowly rolls down the potential giving rise to the present acceleration. Somehow, this is similar to the scenario in the previous subsection; $\hat{\rho}$ plays the role of $\psi$ and its evolution causes symmetry breaking to occur.
The acceleration happens when
\begin{equation}\label{14}
\dot{H}+H^2={1\over 6M_P^2}\left(-2\dot{\phi}^2+2V(\phi)-A(\phi)\hat{\rho}\right)>0.
\end{equation}
Thus, to have an accelerated expansion it is necessary to have $V>\dot{\phi}^2+{A\hat{\rho}\over 2}$. Therefore, if $V(\phi)$ becomes negative for  $\hat{\rho}<\rho_c$ then acceleration does not occur.

\begin{figure}[h]
\centering\epsfig{file=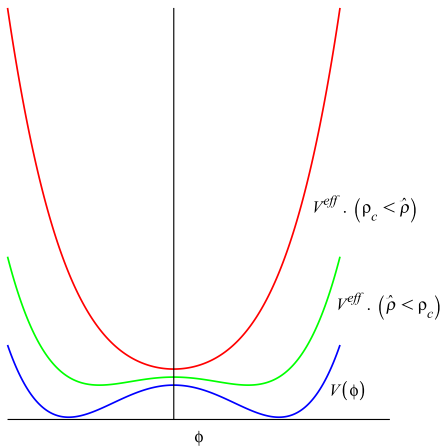,width=6cm,angle=0}
\caption{\footnotesize A schematic illustration of $V(\phi)$  and
$V^{eff}$  in terms of $\phi$
and $\hat{\rho}$. } \label{fig1}
\end{figure}

As an example, take $A(\phi)=1+{\phi^2\over 2M^2}$ and
$V(\phi)=-{\mu^2\over 2}\phi^2+{\lambda\over 4}\phi^4+V_0$, where $M$ and $\mu$ are real constants with mass dimension and  $\lambda$ is a positive dimensionless real number \cite{symmetron}
\begin{equation}\label{15}
V^{eff}={1\over 2}\left({\hat{\rho}\over M^2}-\mu^2\right)\phi^2+\lambda \phi^4+V_0.
\end{equation}
When $\hat{\rho}>M^2\mu^2$ the symmetry is restored and when $\hat{\rho}<M^2\mu^2$  the field rolls down the potential from $\phi=0$, see Fig. \ref{fig1}.

Therefore $V(\phi)$, like the model in the previous subsection, decreases and  can still be positive only when $V_0>0$. But if the field was first at $\phi=0$, it seems that the symmetry breaking is not in favor of  acceleration, see (\ref{3}). A proposal, which was adopted and confirmed via numerical methods in \cite{symmetron} (without a concrete analytical analysis),  is to assume that the initial rapid oscillations of the scalar field about the minimum of the effective potential is converted to  slow roll when $\hat{\rho}<\rho_c$. During this phase, due to presence of the positive term $V_0$, the Universe experiences an accelerated expansion phase.

In the symmetron model proposed in \cite{symmetron}, the symmetry breaking concept was used to explain both the screening effect and the onset of the present positive acceleration.  But the appropriate symmetron mass,  required to explain the screening effect, obtained from gravitational local tests was not consistent with the expected mass of the quintessence without some fine tuning. But from the above discussion, it seems that even though we relax the constraints corresponding to the screening effect and consider only dark sectors, the idea proposed in this model to relate the cosmic acceleration to symmetry breaking, like the hybrid quintessence model, is not adequate.

\section{A new proposal for acceleration from symmetry breaking}

In this part we  present a new proposal for the onset of acceleration where, in contrast to models briefly reviewed in the previous section, the scalar field climbs up the potential when dark matter energy density admits values less than a critical value (this is in favor of acceleration as can be seen from (\ref{3})). As we have shown, a necessary condition for acceleration is the positivity of the potential. We propose a mechanism  in which the positivity of the potential, which is responsible for acceleration, is due to  symmetry breaking. In summary our proposal is modelled as follows:

I- When dark matter density is bigger than a critical value, the scalar field is settled down at the minimum of the U-shaped effective potential, e. g. the stable point $\phi^{*}$, which we take it the same as the minimum of the U-shaped potential satisfying  $V(\phi^{*})\leq 0$. Hence based on (\ref{3}), we have no acceleration during this period of time.

II- When dark matter density becomes less than the critical value, the shape of the effective potential changes.  $\phi^{*}$ becomes the maximum of the effective potential and an unstable point.  $\phi$ acquires a negative squared effective mass and moves from $\phi^{*}$ to the new minimum of the W-shaped effective potential (this is know as the symmetry breaking phenomenon \cite{lind1}).

\begin{figure}[h]
\centering\epsfig{file=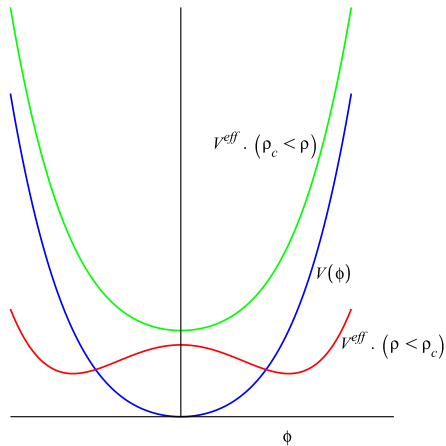,width=6cm,angle=0}
\caption{\footnotesize A schematic illustration of $V(\phi)$,  and
$V^{eff}$  in terms of $\phi$
and $\rho$, in our proposed model for $\phi^*=0$ and $V(\phi=0)=0$.} \label{fig2}
\end{figure}

III- Nonetheless $\phi^{*}$ stills to be the minimum of the potential whose shape does not depend on dark matter density, so while the scalar field descends the effective potential it climbs over its own potential. Therefore we obtain $V(\phi)> 0$ , and the positivity of the potential, causing transition between deceleration and acceleration to occur, is provided. See Fig. \ref{fig2} and compare it with Fig.(\ref{fig1}).

{\it{Hereinafter for simplicity we take the minimum of the potential at $\phi=0$: $\phi^{*}=0$}}. The main difference with Fig. \ref{fig1} is the form of the potential $V$. In Fig. \ref{fig1}, $\phi=0$ is the maximum of the potential while in our model (where the potential is U-shaped), is the minimum. To prevent the confusion between the acceleration due to a positive cosmological constant initially inserted in the potential, as was the case in the models discussed in the previous section (see (\ref{15})), and the acceleration due to the symmetry breaking, we assume $V(\phi=0)\leq 0$.

To realize our plan we need  the scalar field to climb up $V$ while  descending $V^{eff}$.  Hence $V^{eff}_{,\phi}$ and $V_{,\phi}$ must have opposite signs,  i. e.  when the scalar field begins its motion from $\phi=0$, $V^{eff}_{,\phi}<0$ and $V_{,\phi}>0$ (see Fig.(\ref{fig2})). To construct the effective potential we consider an interaction between {\it dark sectors}
\begin{equation}\label{scalar}
\ddot{\phi}+3H\dot{\phi}+V_{,\phi}(\phi)=-h_{,\phi}(\phi,\rho),
\end{equation}
leading to
\begin{equation}\label{17}
V^{eff}_{,\phi}(\phi,\rho)=V_{,\phi}(\phi)+ h_{,\phi}(\phi,\rho),
\end{equation}
where $h$ is an analytical function. We assume that initially, i.e. when $\rho>\rho_c$, $\phi$ is settled down at $\phi=0$ which is the minimum of the effective potential: i.e. $V^{eff}_{,\phi}(\phi=0,\rho>\rho_c)=0$ and $V^{eff}_{,\phi\phi}(\phi=0, \rho>\rho_c)>0$. It is worth to note that it is the effective potential that enters in the equation of motion of the scalar field (\ref{scalar}), while in the Friedmann equation (\ref{3}), it is $V$ that has the key role in the positive acceleration.
When $\rho $ becomes less than the critical value $\rho_c$, the shape of the effective potential changes and $\phi=0$ becomes the maximum of the new effective potential, i.e.  $V^{eff}_{,\phi}(\phi=0,\rho<\rho_c)=0$ and $V^{eff}_{,\phi\phi}(\phi=0, \rho<\rho_c)<0$. $\rho_c$ is defined by $V^{eff}_{,\phi\phi}(\phi=0, \rho=\rho_c)=0$. $\phi=0$ becomes an unstable point, and small perturbations deviate it from zero, so it undergoes a transition towards the new minimum of the effective potential, located at $\phi_{min.}$ defined by $V^{eff}_{,\phi}(\phi=\phi_{min.},\rho<\rho_c)=0$, and $V^{eff}_{,\phi\phi}(\phi=\phi_{min.}, \rho<\rho_c)>0$ \cite{dj}.

To realize this model let us first consider the simple case that $h_{,\phi}(\phi,\rho)$ is a linear function of $\rho$,
\begin{equation}\label{16}
V^{eff}_{,\phi}(\phi,\rho)=V_{,\phi}(\phi)+ B_{,\phi}(\phi)\rho.
\end{equation}
According to our assumptions, when $\rho<\rho_c$ and $\phi$ begins its evolution from $\phi=0$, we must have $\mbox{sgn}\!\left(V^{eff}_{,\phi}V_{,\phi}\right)<0$ for $\rho<\rho_c$.  Hence, as $\rho$ is positive, $B_{,\phi}$ and $V_{,\phi}(\phi)$ must have opposite signs and so a decrease in $\rho$ causes $V^{eff}_{,\phi}(\phi,\rho)$  to have the same sign as $V_{,\phi}(\phi)$ which  contradicts our assumption. Hence the  model (\ref{16}) fails to satisfy our requirements. To elucidate this subject,
let us consider the  potential :
\begin{equation}\label{28}
V={\mu^2\over 2}\phi^2+{\lambda\over 4}\phi^4+V_0;\,\,\lambda>0,\,\, \,\, V_0\leq 0,
\end{equation}
which, in according to our condition I and unlike the potential considered for the symmetron model is U-shaped (the mass term is positive), and $h_{,\phi}=\alpha \phi\hat{\rho}$.
Hence $V^{eff}_{,\phi}=\mu^2\phi+\lambda\phi^3+\alpha \phi \hat{\rho}$. As $\hat{\rho}$ is positive, a positive $\alpha$ does not allow the change of the shape of the effective potential. For $\alpha<0$,  the change of the effective potential from U-shaped to W-shaped occurs when dark matter density increases, which contradicts our condition (II).

Therefore to satisfy (I-III) consistently,  we must try other ansatz where $\rho$ does not appear as a simple linear term in $h_{,\phi}$. In the following we study two examples in this category.

\subsection{A model with transient acceleration}

To obtain a nonlinear interaction  in terms of dark matter density via an action, we use the scalar-tensor action (\ref{9}). We  assume that $A(\phi)$ is just nontrivial, that is $A(\phi)\neq 1$, for dark sectors.  We do not consider ordinary matter and any coupling between dark energy and ordinary baryonic matter (noted by $i=bm$) $A_{i= bm}=1$ (see the discussion after eq.(\ref{9})).

However, as we have seen this
action with a  single scalar field and dark matter does not satisfy our requirements. Hence we assume that $\psi^i$ appearing  in (\ref{9}) consists of a cold
dark matter perfect fluid and a new scalar field $\chi$  with potential $v(\chi)$. We take  $\tilde{g}_{\mu \nu}=A^2(\phi)g_{\mu \nu}$ for dark matter and $\chi$ (we want to relate the cosmic acceleration to the interaction of dark sectors). Our action, after some manipulation, becomes
\begin{eqnarray}\label{18}
S=\int d^4x\sqrt{-g}\left({M_P^2\over 2}R-{1\over 2}g^{\mu \nu}\partial_{\mu}\phi\partial_{\nu}\phi-V(\phi)-{A^2(\phi)\over 2}g^{\mu \nu}\partial_{\mu}\chi\partial_{\nu}\chi-A^4(\phi)v(\chi)\right)+S_m[\tilde{g_{\mu \nu}},\rho].
\end{eqnarray}
The Hubble parameter satisfies the Friedmann equations
\begin{equation}\label{19}
H^2={1\over 3M_P^2}\left({1\over 2}\dot{\phi}^2+{1\over 2}A^2\dot{\chi}^2+V+A^4v+A\hat{\rho}\right),
\end{equation}
\begin{equation}\label{20}
\dot{H}=-{1\over 2M_P^2}\left(\dot{\phi}^2+A^2\dot{\chi}^2+A\hat{\rho}\right),
\end{equation}
where $\rho= A \hat{\rho}$. The equation of motion for the scalar field $\phi$ is
\begin{equation}\label{21}
\ddot{\phi}+3H\dot{\phi}+\left(V_{,\phi}+A_{,\phi}(-A\dot{\chi}^2+4A^3v_{,\chi}+\hat{\rho})\right)=0,
\end{equation}
and $\hat{\rho}$ satisfies $\dot{{\hat{\rho}}}+3H\hat{\rho}=0$, whose solution is given by
\begin{equation}\label{c}
\hat{\rho}(t)=\hat{\rho}_0a^{-3}.
\end{equation}
For $\chi$ we obtain
\begin{equation}\label{22}
\ddot{\chi}+\left(3H+2\left(\dot{\phi}A_{,\phi}\over A\right)\right)\dot{\chi}+A^2v_{,\chi}=0.
\end{equation}

\begin{figure}[H]
\centering\epsfig{file=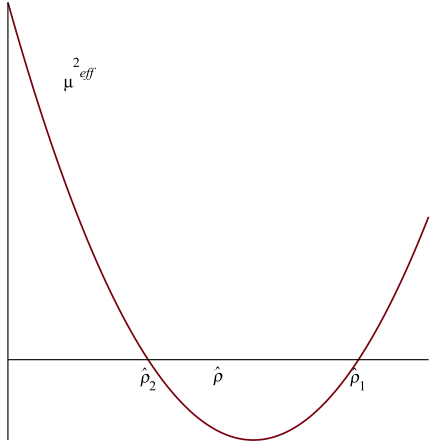,width=6cm,angle=0}
\caption{\footnotesize A schematic illustration of the effective mass squared in terms of ${\hat{\rho}}$}
 \label{fig.5}
\end{figure}

Hereinafter, for the sake of simplicity we ignore $v$, and (\ref{22}) becomes
\begin{equation}\label{23}
\ddot{\chi}+{d\ln(a^3 A^2)\over dt}\dot{\chi}=0,
\end{equation}
whose solution is given by
\begin{equation}\label{24}
\dot{\chi}=C\hat{\rho} A^{-2},
\end{equation}
where $C$ is a numerical constant. Equations (\ref{21}) and (\ref{24}) then lead to
\begin{equation}\label{25}
\ddot{\phi}+3H\dot{\phi}+V_{,\phi}+{C^2\over 4}{\hat{\rho}}^2A^{-4}_{,\phi}+A_{,\phi}{\hat{\rho}}=0,
\end{equation}
so that we may define an effective potential for $\phi$ satisfying
\begin{equation}\label{26}
V^{eff.}(\phi,\hat{\rho})_{,\phi}=V_{,\phi}+A_{,\phi}\hat{\rho}+{C^2\over 4}{\hat{\rho}}^2A^{-4}_{,\phi}.
\end{equation}
Unlike (\ref{16}), $\hat{\rho}$ does not enter in the effective potential, as a simple linear term.  Comparing with (\ref{17}), we find that
\begin{equation}\label{27}
h_{,\phi}(\phi,{\hat{\rho}})= A_{,\phi}{\hat{\rho}}+{C^2\over 4}{\hat{\rho}}^2A^{-4}_{,\phi}.
\end{equation}
The last term is related to the presence of the field $\chi$. Now $V^{eff}_{,\phi}=V_{,\phi}+A_{,\phi}{\hat{\rho}}-C^2A^{-5}A_{,\phi}{\hat{\rho}}^2$ and the signs of $V^{eff}_{,\phi}$ and $V_{,\phi}$ may become opposite, in a consistent way,  by a decreasing ${\hat{\rho}}$.
As a specific example, consider the potential (\ref{28})(note again that the mass term, unlike the symmetron model, is positive)
and $A=A\left({\phi^2\over M^2}\right)$, where $\phi^2\ll M^2$. With this potential form, we have no positive acceleration before the symmetry breaking.  By expanding $A$ we obtain
\begin{eqnarray}\label{29}
&&A\left({\phi^2\over M^2}\right)=A(0)+A'(0){\phi^2\over M^2}+{\cal O}\left({\phi^4\over M^4}\right)\nonumber \\
&&A^{-4}\left({\phi^2\over M^2}\right)=A^{-4}(0)-4{A'(0)\over A^5(0)}{\phi^2\over M^2}+{\cal O}\left({\phi^4\over M^4}\right).
\end{eqnarray}
Therefore (\ref{25}) and (\ref{29}) imply
\begin{equation}\label{30}
\ddot{\phi}+3H\dot{\phi}+\left(-{2C^2A'(0)\over M^2A^5(0)}{\hat{\rho}}^2+{2A'(0)\over M^2}{\hat{\rho}}+\mu^2\right)\phi+\lambda\phi^3=0,
\end{equation}
 where a prime denotes derivative with respect to ${\phi^2\over M^2}$.
The squared effective mass term
\begin{equation}\label{31}
\mu_{eff}^2= -{2C^2A'(0)\over M^2A^5(0)}{\hat{\rho}}^2+{2A'(0)\over M^2}{\hat{\rho}}+\mu^2,
\end{equation}
 is a second order polynomial in terms of ${\hat{\rho}}$ and its sign can change twice at most. We require that for large ${\hat{\rho}}$ the sign is positive and when ${\hat{\rho}}$ is less than a critical value, it becomes negative. Hence we choose ${A'(0)\over A^5(0)}<0$. To allow the sign of the mass term to change, we must have ${A'(0)^2\over M^2}+{2C^2A'(0)\mu^2\over A^5(0)}>0$. With theses conditions it can be shown that (\ref{31}) has two nonnegative roots. We denotes these roots by ${\hat{\rho}}_1$ and ${\hat{\rho}}_2$, where ${\hat{\rho}}_1>{\hat{\rho}}_2$ (see Fig.(\ref{fig.5})).
When ${\hat{\rho}}>{\hat{\rho}}_1$, $\mu_{eff}^2>0$.  The minimum of the effective potential is at $\phi=0$, where $V^{eff}_{,\phi}(\phi=0)=0$ and $V^{eff}_{,\phi\phi}(\phi=0)=\mu^2_{eff}>0$. In this period of time the scalar field is settled at the minimum of the effective potential, $\phi=0$, which is a stable point. We have $V=V_0$ which by definition is non-positive and hence from (\ref{3}) we conclude that there is no positive acceleration. When ${\hat{\rho}}_2<{\hat{\rho}}<{\hat{\rho}}_1$, $\mu_{eff}^2<0$, the effective potential becomes negative with a maximum at $\phi=0$, where $V^{eff}_{,\phi}(\phi=0)=0$ and $V^{eff}_{,\phi\phi}(\phi=0)=\mu^2_{eff}<0$, and two minima at $\phi_{min}^2=-{\mu_{eff}^2\over \lambda}$ where $V^{eff}_{,\phi}(\phi=\phi_{min})=0$ and $V^{eff}_{,\phi\phi}(\phi=\phi_{min})=-2\mu^2_{eff}>0$. In this period of time, $\phi=0$ is an unstable point
and the scalar field rolls down the effective potential from $\phi=0$ towards the new minimum while climbing over its own potential.  Hence both $\phi^2$  and the potential $V(\phi^2)$ increase (see (\ref{28})). Therefore, {\it if the necessary condition for acceleration, i.e. $V>0$, does not hold initially, it may hold afterwards}. This situation continues until ${\hat{\rho}}_2<{\hat{\rho}}$. But when ${\hat{\rho}}<{\hat{\rho}}_2$, the symmetry is restored and $\mu_{eff}^2>0$. In this case we have only a minimum at $\phi=0$ and the system returns to its initial state.

Let us now be more specific and as an example take  $A(\phi)=\left(1+{\phi^2\over M^2}\right)^{-1}$ where
$M$, as before, is a mass scale in the theory such that $\phi^2\ll M^2$. Therefore we can expand $A({\phi^2\over M^2})$ as
\begin{equation}\label{d1}
A({\phi^2\over M^2})=1-{\phi^2\over M^2}+\mathcal{O}\left({\phi^4\over M^4}\right)
\end{equation}
In this way, the second term in the expansion, modifies the mass term of the scalar field. Note that $A(\phi)$ in our example differs from \cite{symmetron} by a minus sign. Dark matter densities for $\mu_{eff}=0$ are given by
\begin{eqnarray}\label{d2}
\hat{\rho_1}&=&{1+\sqrt{1-2C^2M^2\mu^2}\over 2C^2}\nonumber \\
\hat{\rho_2}&=&{1-\sqrt{1-2C^2M^2\mu^2}\over 2C^2}.
\end{eqnarray}
By assuming that initially $\phi=0$ and from the fact that the dark energy and the dark matter densities have the same order of magnitude, we find the order of $\hat{\rho_1}$ as
$\hat{\rho_1}\sim 3M_P^2 H_0^2$ \cite{rep}, where $H_0$ is the present Hubble parameter. Also, expecting that the acceleration lasts at least for a Hubble time gives
${\dot{\hat{\rho}}\over H_0}\simeq \Delta \hat{\rho} \sim H_0^2 M_P^2$ \cite{rep}. But $\Delta \hat{\rho}={\sqrt{1-2C^2M^2\mu^2}\over C^2}$. These imply ${1\over 2C^2}\sim  M_P^2 H_0^2$.
The solution (\ref{d2}) is valid for  $\mu^2< {1\over 2C^2M^2}$, therefore
\begin{equation}\label{d4}
{\mu^2\over H_0^2}\precsim {M_P^2\over M^2}.
\end{equation}

 In \cite{symmetron},\cite{khoury}, where unlike our model, the quintessence besides the dark matter is also coupled to ordinary baryonic mater, the squared of the effective mass is linear in terms of $\hat{\rho}$, therefore its signs changes only when the matter density crosses $\hat{\rho}= \mu^2 M^2$. This together with $\hat{\rho}\sim 3M_P^2 H_0^2$ \cite{symmetron}, obtained from the Friedmann equation, results in ${\mu^2\over H_0^2}\sim{ M_P^2\over M^2}$. If one considers $M^2\ll M_P^2$, obtained from local screening test of gravity \cite{symmetron}, he derives a large mass for the quintessence, i.e. $\mu^2 \gg H_0^2$. This  forces the quintessence to overshoot rapidly and to oscillate about the minimum of the effective potential stopping the acceleration. In our model the condition on the quintessence mass, differs from \cite{symmetron}, as it can be seen from (\ref{d4}). Besides we have not considered any coupling between quintessence and baryonic matter (as emphasized in subsection 3.2 and 4.1). However if one sets a conformal coupling between quintessence and ordinary matter, variation of the action (\ref{18}) with respect to $\phi$ and potential-less $\chi$ gives
\begin{eqnarray}\label{d3}
&&\Box \phi+V_{,\phi}+A A_{,\phi} \nabla \chi.\nabla \chi+A^{-1}A_{,\phi}\rho+{A_{bm}}^{-1}A_{bm,\phi}\rho_{bm}=0 \nonumber \\
&&\nabla.\left(A^2\nabla\chi\right)=0
\end{eqnarray}
where $A_{bm}$ is the conformal coupling to ordinary matter: $\rho_{bm}$ . If we set $A_{bm}=1$, as before,  there is no coupling between quintessence and ordinary matter.
Expanding $\phi$ about the background, $\phi(t)$ as: $\phi=\phi(t)+\varphi(x)$, and by considering the equation of motion for $\varphi$, one can obtain the static potential due to the coupling $A_{bm}$ and compare it with the gravitational Newtonian potential. It is expected that this new potential will be negligible in local test of gravity due to the screening effect \cite{symmetron}.  In the absence of $\chi$, Newtonian and post Newtonian approximation  put constraints on  $M$ as $M\ll 10^{-4}M_P$. The presence of the field $\chi$ modifies $\varphi$ equation. So as an outlook it may be interesting to see whether a theory including new scalar fields is able to modify the constraint on $M$ as was suggested in \cite{prl}. Modification of $M$ in a hybrid model in modified theory of gravity was also discussed in \cite{Bamba}.

At the end of this part let us illustrate our results by specifying the potential. As an example, we take the potential as $V(\phi)={\lambda\over 4}\phi^4$ with $\lambda>0$ and take $A$ as (\ref{d1})(note that the symmetry breaking is not allowed with this potential in \cite{symmetron}, where only a single field is employed). Equation (\ref{26}) reduces to $V^{eff}={\lambda \over 4}\phi^4+\left(1+{\phi^2\over M^2}\right)^{-1}\hat{\rho}+{C^2\over 4}{\hat{\rho}}^2\left(1+{\phi^2\over M^2}\right)^{4}$. It is clear that $\phi=0$ is the minimum of $V(\phi)={\lambda\over 4}\phi^4$. We  also have $V^{eff}_{,\phi}(\phi=0)=0$ and   $\mu_{eff}^2=V^{eff}_{,\phi\phi}(\phi=0)={2C^2\over M^2}{\hat{\rho}}^2-{2\over M^2}\hat{\rho}$. When $\hat{\rho}>{1\over C^2}$, the effective potential has only a minimum at $\phi=0$.  So we assume that $\phi$ resides initially at the stable point $\phi=0$. When $\hat{\rho}$ decreases such that  $\hat{\rho}<{1\over C^2}$, the effective potential becomes W-shaped and gains two minima at $\phi=\pm{\sqrt{2\lambda\rho(1-C^2\rho^2)}\over {\lambda M}}$ and a maximum at $\phi=0$. Hence when $\hat{\rho}<{1\over C^2}$, $\phi=0$ becomes an unstable point and the scalar field rolls down to the new minimum
of the effective potential. But as $\phi$ moves from $\phi=0$, $\phi^2$ increases and $V^{eff}={\lambda \over 4}\phi^4$ increases too and becomes positive which is the necessary condition for the onset of  acceleration.

\begin{figure}[H]
\centering\epsfig{file=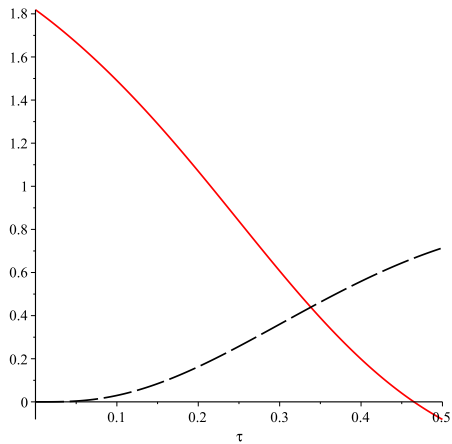,width=6cm,angle=0}
\caption{\footnotesize The potential (dashed line) and the deceleration factor (solid line) in terms of dimensionless time $\tau$ for $\tilde{M}=100,\,\, \tilde{C}=0.25, \,\, \tilde{\lambda}=1, \,\,\varphi(0)=0,\,\, \tilde{{\hat{\rho}}}(0.5)=0.3, \,\,\varphi(0.5)=1.3, \,\,\, \tilde{H}(0.5)=1$} \label{fig.3}
\end{figure}

The deceleration parameter and potential, for this example, are numerically shown  in terms of the dimensionless time $\tau=H_0t$ in Fig. \ref{fig.3}, showing a deceleration to acceleration phase transition when the field climbs over  its own potential ($H_0$ is the present time Hubble parameter).
For this numerical illustrations we have introduced dimensionless parameters
 $\tilde{\lambda}={M_P^2\over H_0^2}\lambda,\, \tilde{{\hat{\rho}}}={{\hat{\rho}}\over M_P^2H_0^2}, \,\varphi={\phi\over M_p},\, \tilde{C}^2=H_0^2M_P^2 C^2, \, \tilde{\mu}={\mu \over H_0},\, \tilde{M}={M\over M_p},\, \tau=tH_0,\, \tilde{H}={H\over H_0}$.  At $t=0$, the potential vanishes and the universe is in a deceleration phase. When the squared effective mass becomes negative, the potential increases and a positive acceleration sets in.

 It is worth noting that the acceleration in this model is transient; as ${\hat{\rho}}$ reduces more and becomes equal to or less than the second root of (\ref{31}), the symmetry is restored and ultimately the effective potential takes the same shape as the initial potential. So the scalar field rolls down the effective potential as well as its own potential and will settle down to its initial value $\phi=0$, its own potential becomes $V_0$ which by definition is not positive and acceleration will  end.

\subsection{A model with persistent acceleration}

To shed light on how a persistent acceleration can be obtained via symmetry breaking, we take advantage of interaction between dark sectors such that  when $\rho$ becomes less than a critical value, the sign of the scalar field effective mass changes. Here we abandon our previous approach and do not insist on  the precise form of the interaction via metric modification in the dark matter sector.  We work in the Einstein frame and $\rho$ denotes the usual dark matter energy density in the Einstein frame.

We require that the interaction behaves as a mass term and its sign changes during $\rho$ evolution. The equations of motions for the scalar field dark energy interacting with dark matter via a source $f(\rho)B(\phi)$, where $f$ and $B$ are analytical functions (this can be considered as a generalization of interaction $B(\phi)\rho$ considered in the literature \cite{faro}  and is a special case of (\ref{17})) are
\begin{eqnarray}\label{32}
&&\ddot{\phi}+3H\dot{\phi}+V_{,\phi}(\phi)= -f(\rho)B_{,\phi}(\phi), \nonumber \\
&&\dot{\rho}+3H\rho=f(\rho)B_{,\phi}(\phi)\dot{\phi}.
\end{eqnarray}
The effective potential satisfies $V^{eff}_{,\phi}=V_{,\phi}(\phi)+f(\rho)B_{,\phi}(\phi)$ .

Our plan is as follows: we require $f(\rho)B(\phi)$ to behave as a mass-like term. We choose $f(\rho)$ as an analytic function whose sign changes when $\rho$ decreases.
We assume that when $\rho>\rho_c$, the effective potential has a minimum at a constant $\phi^*$ which is the true vacuum of the system. At this point $V(\phi^*)\leq 0$ and so we have a deceleration phase. When the scalar field is settled down to this vacuum, $\rho$ continues to decrease. However, when $\rho<\rho_c$, the symmetry is broken and $\phi^*$ is no more the true vacuum and the scalar field rolls down the effective potential so $\phi^2$ increases while $\rho$ continues to decrease.  The increasing $\phi^2$ makes the potential to become greater than zero which is necessary for acceleration. The scalar field eventually lies at a fixed point $\phi_c$ (the new minimum of the effective potential) whose  own potential is positive, giving rise to a permanent acceleration; the Universe tends to a de Sitter space-time with a constant Hubble parameter
 \begin{equation}\label{33}
 H^2={1\over 3M_P^2}V(\phi_c).
 \end{equation}
To study the evolution and stability of this model let us write the equations of motion in the form of an autonomous system
\begin{eqnarray}\label{34}
\dot{u}&=&-3Hu-V_{,\phi}-f(\rho)B_{,\phi},\nonumber \\
\dot{\rho}&=&-3H\rho+f(\rho)B_{,\phi}u,\nonumber \\
\dot{H}&=&-{1\over 2M_P^2}\left(u^2+\rho\right),\nonumber \\
\dot{\phi}&=&u.
\end{eqnarray}
The critical points $\bar{\phi}, \bar{\rho}$ of this system are given by
\begin{equation}\label{35}
\bar{u}=\dot{\phi}|_{\bar{\phi}}=0,\,\, \bar{\rho}=0,\,\, (V_{,\phi}+f(\rho)B_{,\phi})|_{\rho=0, \phi=\bar{\phi}}=0.
\end{equation}
We consider small homogeneous variations about the critical points : $\delta \phi,\,\, \delta u ,\,\, \delta H,\,\, \delta \rho$ in (\ref{34}) to obtain

\begin{equation}\label{36}
{d\over d{t}}{ \left( \begin{array}{cccc}
\delta u \\
\delta \rho \\
\delta H \\
\delta \phi
\end{array} \right)}=\mathcal{M} \left( \begin{array}{cccc}
\delta u \\
\delta \rho \\
\delta H \\
\delta \phi
\end{array} \right),
\end{equation}
where
\begin{equation}\label{37}
\mathcal{M}=\left( \begin {array}{cccc} -3\,H& -\left( {\frac {\rm d}{
{\rm d}\rho}}f \right) {\frac {\rm d}{{\rm d}\phi
}}B &-3\,u&-{\frac {{\rm d}^{2}}{{\rm d}{\phi}^{2}
}}V  -f {\frac {{\rm d}^{2}}{
{\rm d}{\phi}^{2}}}B  \\ \noalign{\medskip}f
  {\frac {\rm d}{{\rm d}\phi}}B  &-3\,H+ \left( {\frac {\rm d}{{\rm d}\rho}}f   \right)  \left( {\frac {\rm d}{{\rm d}\phi}}B   \right) u&-3\,\rho& f   \left( {
\frac {{\rm d}^{2}}{{\rm d}{\phi}^{2}}}B   \right)
u\\ \noalign{\medskip}-{\frac {u}{{M_P}^{2}}}&-{1\over 2{M_P}^{2}}&0&0
\\ \noalign{\medskip}1&0&0&0\end {array} \right).
\end{equation}
The roots of the characteristic polynomial of $\mathcal{M}$ are a simple $x=0$, and $x$ satisfies
\begin{equation}\label{38}
x^3+6Hx^2+\left(ff_{,\rho}B_{,\phi}^2+9H^2+(V_{,\phi\phi}+fB_{,\phi\phi})\right)x+
3H\left(V_{,\phi\phi}+fB_{,\phi\phi}\right)=0,
\end{equation}
evaluated at the critical points. The corresponding Routh table is
\begin{equation}\label{39}
\left[ \begin{array}{ccc}
1 &
9H^2+B_{,\phi}^2ff_{,\rho}+V_{,\phi\phi}+fB_{,\phi\phi} &
x^3 \\
6H & 3H(B_{,\phi\phi}f+V_{,\phi \phi}) & x^2 \\
9H^2+B_{,\phi}^2ff_{,\rho}+{1\over 2}(V_{,\phi\phi}+fB_{,\phi\phi}) &0 &x \\
3H(B_{,\phi\phi}f+V_{,\phi \phi})& 0 & 1
\end{array} \right].
\end{equation}
So the system is stable at a critical point determined by (\ref{35}) provided that $V^{eff}_{,\phi \phi}=B_{,\phi\phi}f+V_{,\phi \phi}>0$, and  $9H^2+B_{,\phi}^2ff_{,\rho}+{1\over 2}(V_{,\phi\phi}+fB_{,\phi\phi})>0$ at this point.

As an example we choose the potential as $V={1\over 2}\mu^2\phi^2+{\lambda\over 4}\phi^4$, $\lambda>0$. So in order that $f(\rho)B(\phi)$ behaves as a mass-like term we take $B(\phi)\propto \phi^2$. As was mentioned  before we choose $f(\rho)$ as an analytic function such that the effective mass term sign changes when $\rho$ decreases. Our simplest choice is then $B(\phi)f(\rho)={\phi^2\over 2M^2}(\rho-\rho_c)$, where $M$ and $\rho_c$ are two constants with dimension of $[mass]$ and $[mass]^4$ respectively and we take $\rho_c>\mu^2 M^2$.
Now let us study the behavior of the system. When $\rho>\rho_c-\mu^2 M^2$, the effective potential has a minimum at $\phi=0$ and $\phi=0$ is a solution of the system (\ref{32}). During the time when $\phi=0$, $\rho$ continues to decrease. When $\rho<\rho_c-\mu^2 M^2$, $\phi=0$ is longer the true vacuum and the scalar field rolls down the effective potential so $\phi^2$ increases.  By increasing  $\phi^2$ the potential becomes greater than zero which is necessary for acceleration. Eventually $\phi^2$ settles at $ \phi^2={\rho_c-\mu^2 M^2\over \lambda M^2}$ (the new minimum of the effective potential), while $\rho\to 0$. Therefore the Universe tends to a de Sitter space-time
 with a constant Hubble parameter
 \begin{equation}\label{40}
 H^2={1\over 12M_P^2}{\rho_c^2-\mu^4 M^4\over \lambda M^4}.
 \end{equation}

\begin{figure}[H]
\centering\epsfig{file=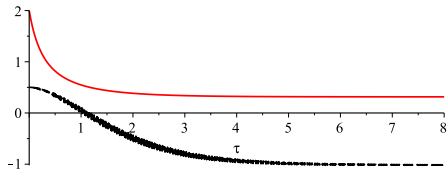,width=6cm,angle=0}
\caption{\footnotesize The deceleration (dashed) and Hubble parameters (line) in terms of dimensionless time $\tau$ for $\tilde {\rho}(0)=12, \mu=0, \tilde{\lambda}=10^{10}, \tilde{\rho}_c=11, \varphi(0)=0, \tilde{H}(0)=2$.} \label{fig4}
\end{figure}

As an illustration, let us use the equations of motions rewritten in terms of dimensionless parameters defined previously
\begin{eqnarray}\label{41}
&&{d^2\varphi\over d\tau^2}+3\tilde{H}{d\varphi\over d\tau}+\tilde {\mu^2} \varphi +\tilde{\lambda}\varphi^3=-{1\over \tilde{M}^2}(\tilde {\rho}-\tilde{\rho_c})\varphi,\nonumber \\
&&{d\tilde{\rho}\over d\tau}+3\tilde{H}\tilde{\rho}={1\over \tilde{M}^2}(\tilde {\rho}-\tilde{\rho_c})\varphi{d\varphi \over d\tau},\nonumber \\
&& {d\tilde{H}\over d\tau}=-{1\over 2}\left(\left({d\varphi\over d\tau}\right)^2+\tilde{\rho}\right),
\end{eqnarray}
 to show the deceleration parameter in terms of dimensionless time. The deceleration and  Hubble parameter are shown in Fig. \ref{fig4} for $\{\tilde {\rho}(0)=12, \mu=0, \tilde{\lambda}=10^{10}, \tilde{\rho}_c=11, \tilde{M}=0.01, \varphi(0)=0, \tilde{H}(0)=2\}$, showing that the Universe enters an acceleration phase and finally becomes a de Sitter space-time.

\section{Summary}
We have briefly studied and reviewed the possible relation between the symmetry breaking and the onset of cosmic acceleration in two models proposed in the literature, i.e. hybrid and symmetron scalar field dark energy models. We delineated the role of the positive scalar field potential which drives the acceleration and showed that in these  models the symmetry breaking is not in favor of the positive acceleration. In section four, we tried to relate the positivity of the potential to the spontaneous symmetry breaking of the effective potential constructed from the interaction between dark sectors. To do so, we required that the scalar field climbs over its own potential while moving down along the effective potential. We proved that this requirement cannot be satisfied in a model whose interaction is linear in terms of the matter density (like the symmetron model). To obtain a non linear interaction,  we introduced and additional scalar field in the dark energy sector in a scalar tensor type action consisting only of dark species. For the sake of simplicity we took it without a
potential but one may examine a general potential for the problem. It was shown and illustrated that in this framework a transient deceleration to acceleration phase
transition may occur. In the second model,  to obtain a permanent acceleration, we presented an interaction between the dark sector (which does
not have its roots in a fundamental action). We showed that in this new model, the Universe may evolve from a deceleration phase to a de Sitter
space-time through  symmetry breaking. We verified the stability of the model and illustrated our results through a numerical example.

\end{document}